\documentclass[a4paper]{PoS}

\usepackage{amsmath, amssymb, amsfonts}
\usepackage{graphicx}

\def\beq{\begin{equation}}
\def\eeq{\end{equation}}
\def\beqa{\begin{eqnarray}}
\def\eeqa{\end{eqnarray}}
\def\ben{\begin{enumerate}}
\def\een{\end{enumerate}}
\def\bit{\begin{itemize}}
\def\eit{\end{itemize}}

\def\sY{{\mathsf{Y}}}

\def\nn{\nonumber}

\title{$SO(10)$ inspired extended GMSB models}

\ShortTitle{$SO(10)$ inspired extended GMSB models}

\author{\speaker{Tomasz Jeli\'nski}
        \\
        Institute of Physics, University of Silesia,\\
        Uniwersytecka 4, PL-40-007 Katowice, Poland\\
        E-mail: \email{tomasz.jelinski@us.edu.pl}}

\abstract{
Influence of messenger-matter superpotential interactions on the renormalization of Yukawa couplings in the context of extended GMSB models is analysed.  We  present a convenient method for treating decoupling of messengers and related redefinition of MSSM fields. 
Discussed approach is 
used to study 
top-bottom-tau Yukawa unification within specific $SO(10)$ inspired GUT model. 
}

\FullConference{Proceedings of the Corfu Summer Institute 2014 "School and Workshops on Elementary Particle Physics and Gravity",\\
		3-21 September 2014\\
		Corfu, Greece }

\begin{document}

\section{Introduction}\label{sec1}

Among supersymmetric GUT models those which involve
%are based on 
the Gauge Mediated Supersymmetry Breaking (GMSB) 
mechanism  
\cite{Giudice:1998bp} are of great interest for a long time. They exhibit many desired features, 
but they also require
%need 
rather large masses of sparticles to accommodate for 125 GeV Higgs mass \cite{Draper:2011aa}. Recently, it was shown that this issue can be naturally addressed  in so-called extended GMSB (EGMSB) models, in which messengers interact with MSSM fields through superpotential couplings \cite{Dine:1996xk,Giudice:1997ni}. 
%{\bf 
%
Such couplings are natural because messengers carry the same charges under $G_{SM}=SU(3)\times SU(2)\times U(1)$ as MSSM fields hence they can interact  through superpotential in an analogous way.
%} 
These couplings 
%where 
generate 1-loop $A$-terms (and 1- and 2-loop soft masses) at the messenger scale $M$ \cite{Dine:1996xk,Giudice:1997ni,Chacko:2001km,Joaquim:2006uz,Joaquim:2006mn,Evans:2013kxa,Jelinski:2013kta}. 
%are generated at the messenger scale $M$. 
In consequence, in many realizations of the EGMSB models stop mixing at the electroweak symmetry breaking (EWSB) scale is enhanced what in turn helps  to get 
%accomodate for 
proper Higgs mass \cite{Evans:2011bea,Jelinski:2011xe,Kang:2012ra,Albaid:2012qk}. The other motivation for considering these models is that they provide a reasonable
%/controlable/interesting tool/well-behaved-understood
framework to explore non-minimal flavour violation scenarios \cite{Albaid:2012qk,Shadmi:2011hs,Abdullah:2012tq,Calibbi:2013mka,Galon:2013jba, Jelinski:2014uba}, e.g. those which are inspired by the F-theory geometrical constructions \cite{Yukawa-HV1,HV-E8,Pawelczyk:2010xh} or Froggatt-Nielsen type models \cite{Froggatt-Nielsen}. As a consequence of extra contributions to soft masses, EGMSB models have also quite rich and interesting 
%low-energy 
phenomenology, where 
%including 
non-standard NLSP/NNLSP patterns like stop/bino or sneutrino/stau \cite{Jelinski:2011xe} can be realized. 
%/arise
%new interesting scenarios arise

It is clear 
%well-known 
that in the GMSB models the presence of messengers changes running of gauge couplings $g_r$ above messenger scale $M$  with respect to the standard RGE evolution of MSSM parameters.  
It is well known that 
when messengers are in full representations of GUT gauge group then unification of $g_r$ is not spoiled by their presence \cite{Giudice:1998bp}. Moreover, via RG equations, they also indirectly change running of top, bottom and tau Yukawa couplings $y_{t,b,\tau}$ \cite{Bagger:1996ei}. Hence, it is natural to raise a question about 
Yukawa couplings 
%their 
unification  when one allows for superpotential couplings between messengers and matter, as in EGMSB models. That issue can be rephrased also in the following form. Assuming unification of Yukawa couplings $y_{t,b,\tau}$ at  the GUT scale $M_{GUT}\approx2\times 10^{16}\,\mathrm{GeV}$, one can ask what are their values at the EWSB scale and do they match values derived from fermion masses \cite{Pierce:1996zz}. 

We show that the 
%wave-function renormalization 
method \cite{Evans:2013kxa} which was developed to properly derive soft terms in EGMSB models 
%when messenger-matter couplings are present 
can be used
%/[dostosowana] 
to get information about the running of Yukawa couplings in those scenarios. It it especially useful in the top-down approach where one assumes precise Yukawa couplings unification at the GUT scale $M_{GUT}$ and derives their values at the EWSB scale from RGE evolution. 
%Then the 
That method allows to keep track of
%/follow/control/? 
messenger-matter kinetic mixing and study how 
%the presence of 
messengers superpotential couplings contribute to values of Yukawas at the EWSB scale. 
We apply our method to analyse the issue of top-bottom-tau unification in the context of specific $SO(10)$ inspired GUT model with minimal matter content and  only one messenger-matter superpotential coupling. Such $SO(10)$-type GUT model, see e.g. \cite{Blazek:2002ta,Raby:2003in,Badziak:2011wm}, provides a natural setup to present the method and expose issues related to the kinetic mixing of messengers and MSSM matter triggered by superpotential couplings. 

\section{Extended GMSB models}\label{sec2}
Let us briefly recall the salient features of so-called extended GMSB models. For more details see e.g. \cite{Dine:1996xk,Giudice:1997ni, Chacko:2001km,Joaquim:2006uz,Evans:2013kxa,Jelinski:2013kta,Byakti:2013ti}. As in the standard GMSB models \cite{Giudice:1998bp} supersymmetry is broken in the hidden sector by the $F$-term of the spurion superfield $X$. Messenger sector consists of chiral superfields ${\sY}_A=(Y_A,\overline{Y}_A)$ in vectorial representations of $G_{SM}$. We assume that all the messenger fields couple in the same way to the spurion $X$ through a superpotential term $XY_A\overline{Y}_A$ \cite{Craig:2012xp}. When $X$ gets vev $\left<X\right>=M+\theta^2F_X$ that coupling gives mass $M$ to the messengers and induces additional SUSY breaking masses $F_XY\overline{Y}$ for their scalar components. For simplicity, let us consider messenger fields in the following representations of $G_{SM}$:
\beqa\label{messlist}
&&Y_{H_u}: (1,2)_{1/2},\quad Y_Q: (3,2)_{1/6},\quad Y_{\overline{U}}: (\overline{3},1)_{-2/3},\nonumber\\[-0.5\baselineskip]
\\[-0.5\baselineskip]
&&Y_{\overline{D}}: (\overline{3},1)_{1/3},\quad Y_L: (1,2)_{-1/2}, \quad Y_{\overline{E}}: (1,1)_{1}\nonumber
\eeqa
and their partners in conjugate representations: $Y_{\overline{H}_u} (1,2)_{-1/2}$, $Y_{\overline{Q}} (\overline{3},2)_{-1/6}$, $Y_{\overline{U}}(3,1)_{2/3}$ etc. We have not included $Y_{H_d}$  in the list \eqref{messlist} as it would have the same quantum numbers as $Y_{L}$. Moreover, we also allow for a singlet messenger $Y_{N_R} (1,1)_{0}$ under $G_{SM}$. Our choice of messengers  
is motivated by most 
%popular/
common constructions of GUT models based on $SU(5)$ and $SO(10)$, where after breaking of $G_{GUT}$ to  $G_{SM}$ all light fields are only in representations shown in \eqref{messlist}. Let us note that
%/Of course 
more complicated setups are also possible e.g. when messengers are in higher-dimensional representations of $G_{SM}$. For example, when spectrum contains messenger $Y_{S}$ in $(\overline{6},1)_{-1/3}$ of $G_{SM}$ then the following coupling is possible $QY_SQ$. 

In the EGMSB models messengers ${\sY}_A$ couple to MSSM fields $\Phi_a$ not only by gauge fields but also directly through the following superpotential:
\beqa\label{W3}
W_3&=&\frac{1}{6}\lambda_{ijk}\Phi_i\Phi_j\Phi_k\nonumber\\
&=&\frac{1}{6}y_{abc}\Phi_a\Phi_b\Phi_c+\frac{1}{2}h_{abC}\Phi_a\Phi_b\sY_C+\frac{1}{2}h_{aBC}\Phi_a\sY_B\sY_C+\frac{1}{6}\eta_{ABC}\sY_A\sY_B\sY_C,
\eeqa
what leads to generating 1-loop $A$-terms and 1- and 2-loop soft masses \cite{Dine:1996xk,Giudice:1997ni,Chacko:2001km,Evans:2013kxa,Jelinski:2013kta}. We shall focus on the regime $F/M^2\ll 1$ in which 2-loop soft masses dominate over 1-loop soft masses \cite{Dine:1996xk,Giudice:1997ni}. Let us note that when $\lambda_{ijk}$ couplings are zero then one gets the  standard GMSB model. For discussion of the phenomenology of the EGMSB models see e.g. \cite{Evans:2011bea,Jelinski:2011xe,Kang:2012ra,Albaid:2012qk}. 

\section{Decoupling and redefinition of fields}\label{sec3}

It is clear that in the presence of superpotential couplings \eqref{W3} between messengers and matter both fields are subject to kinetic mixing. Let us note that in the models under consideration there are two types of such mixing: (i) between two fields ($H_u\leftrightarrow Y_{H_u}$, $Q\leftrightarrow Y_Q$, $U\leftrightarrow Y_U$), and  (ii) between three fields ($H_d\leftrightarrow L\leftrightarrow Y_{L}$) when down-type Higgs field $H_d$ mixes with $L$ and messenger field $Y_L$. 
%We shall focus on the first case. Generalization to mixing $$ is straightforward. 
For example, messenger field $Y_{Q}$ with the same quantum numbers as quark field $Q$ can mix with the latter when superpotential contains $H_uY_QU$ or $H_dY_QD$.  
%{\bf [Is it the only possibility?]}. 
As a consequence of kinetic mixing, wave-function renormalization $Z_{ij}$ receives non-diagonal loop corrections, even if it was chosen to be diagonal at the GUT scale $M_{GUT}$. That in turn, according to the non-renormalization theorem \cite{Grisaru:1979wc}, influences running of Yukawa couplings $\widetilde{\lambda}_{ijk}$ between canonically normalized fields $\widetilde{\Phi}_i$. A method presented in \cite{Evans:2013kxa}, originally developed to properly compute 2-loop soft-masses induced by \eqref{W3}, is especially convenient to reveal influence of messenger-matter interactions on Yukawa couplings $\widetilde{y}_{abc}$ renormalization. At the scale\footnote{$\mu$ is a renormalization scale, while $Q$ is an arbitrary scale which we set to $1\,\mathrm{GeV}$.} $t=\ln \mu/Q$, lower than the GUT scale $t_{GUT}=\ln M_{GUT}/Q$, the model is characterized by the superpotential and the K\"ahler potential: 
\beqa\label{Wlam}
W=\frac{1}{6}\lambda_{ijk}\Phi_i\Phi_j\Phi_k+\frac{1}{2}M_{ij}\Phi_i\Phi_j,\quad K=\Phi_i^{\dag}Z_{ij}(t)\Phi_j,
\eeqa
where $Z_{ij}$ is positive-definite Hermitian wave-function renormalization matrix and $Z(t_{GUT})_{ij}=\delta_{ij}$, while $M_{ij}$ is complex symmetric mass matrix. $M_{ij}$ is chosen such that at scale $t_{GUT}$ messenger fields $\sY_A$ are massive i.e. they have mass terms of the form $MY_A\overline{Y}_A$ while  MSSM fields $\Phi_a$ are light.\footnote{'Massive' means that fields $\sY_A$ have mass of the order of $M$, while 'light' means that fields $\Phi_a$ are massless or have a supersymmetric mass of the order of EWSB scale.} In the so-called holomorphic scheme \cite{Evans:2013kxa}, in which K\"ahler potential is not canonical  \eqref{Wlam}, couplings $\lambda_{ijk}$ do not run. On the other hand,  
% Light is better description because Higgses has to be massive even before gauge symmetry breaking
%
it is clear that the running couplings $\widetilde{\lambda}_{ijk}(t)$ and masses $\widetilde{M}_{ij}(t)$ of canonically normalized fields $\widetilde{\Phi}_i=Z^{-1/2}_{ij}\Phi_j$ are related to their holomorphic counterparts, $\lambda_{ijk}$ and $M_{ij}$,  
%quantities 
in the following way:
\beqa\label{runpar}
\widetilde{\lambda}_{ijk}(t)=\lambda_{i'j'k'}Z^{-1/2}_{i'i}Z^{-1/2}_{j'j}Z^{-1/2}_{k'k}, \quad \widetilde{M}_{ij}(t)=M_{i'j'}Z^{-1/2}_{i'i}Z^{-1/2}_{j'j}.
\eeqa
Hence non-diagonal $Z_{ij}$, generated by loop corrections, induce mixing mass terms between messenger fields and MSSM fields. In other words: RGE evolution of $Z_{ij}$ reintroduces mixing mass terms between both fields. Hence to consistently define theory below messenger scale $\widetilde{M}(t_M)=M$, one has to recognize light states (matter fields) and heavy states (messengers), properly decouple the latter at $M$, and appropriately derive couplings between light states (physical Yukawa couplings). To do that, one has to know which combination of original fields $\Phi_i$ are massive and which are light. One way to do this is to, first, diagonalize $Z_{ij}$ and then find null vectors\footnote{Or, more generally, light states.} of mass matrix $\widetilde{M}$ \eqref{runpar} and orthogonal combinations (heavy states). 
%
%\footnote{It is clear that $\widetilde{M}$ has to have precisely one null vector in representation $\overline{5}$ of $SU(5)$ and precisely one in $10$. Morover the %spectrum has to contain two fields in representations $5$ and $\overline{5}$ whose doublet components have mass of the order of the EWSB scale and triplets components have mass of order $M_{GUT}$. Otherwise the matter content of the model does not match that of the MSSM.} 
%
The other method, which seems to be more convenient in the discussed case, is to use Cholesky decomposition \cite{Horn} of $Z_{ij}$, which at once allows to find such a combination of holomorphic fields which stays heavy or massless under loop corrections. Here, instead of computing $Z^{-1/2}$ and then finding null vectors of $\widetilde{M}$ and their orthogonal combinations, one defines physical fields in the following way: 
\beqa\label{mix2}
%Z=V^\dag V,\qquad
 \widetilde{\Phi}=\left(\begin{array}{c}\widetilde{\phi}\\\widetilde{Y}\end{array}\right)=V\left(\begin{array}{c}\phi\\Y\end{array}\right)=\left(\begin{array}{cc}V_{11}&V_{12}\\0&V_{22}\end{array}\right)\left(\begin{array}{c}\phi\\Y\end{array}\right),
\eeqa
where $V$ is an upper triangular matrix which enters Cholesky decomposition of $Z$:
\beq 
Z=V^\dag V.
\eeq
In \eqref{mix2} we displayed an example of such redefinition when only two fields, $\phi$ and $Y$, are subject to kinetic mixing $\phi\leftrightarrow Y$. Generalization to mixing of three or more fields is straightforward. 
%
%{\bf It is easy to see that such redefinition of fields respects the form of mass matrix i.e. it does not change null vectors of the mass matrix.}
%
As a consequence, the running physical superpotential couplings $\widetilde{\lambda}_{ijk}(t)$ can be now written as:
\beq\label{lat}
\widetilde{\lambda}_{ijk}(t)=\lambda_{i'j'k'}V^{-1}_{i'i}V^{-1}_{j'j}V^{-1}_{k'k}.
\eeq
Let us stress that the advantage of such approach is that, now, running couplings $\widetilde{\lambda}_{abc}(t)$ between light fields $\widetilde{\Phi}_a$, i.e. physical Yukawa couplings, are expressed only in terms of wave-function renormalization without the need for finding null vectors of the mass matrix $\widetilde{M}$. Moreover, using \eqref{lat}  one can show that  $\widetilde{\lambda}_{abc}(t)$ are completely determined  by the $Z_{ij}$ running:
\beq\label{lambdaphys}
\widetilde{\lambda}_{abc}(t)=\frac{\lambda_{abc}}{\sqrt{{Z}_{aa}(t){Z}_{bb}(t){Z}_{cc}(t)}},
\eeq
with no sum over repeated indices. It is known \cite{Evans:2013kxa,Martin:1993zk} that $Z_{ij}$ evolves according to the following one-loop renormalization group equation:
\beq\label{RGEZ}
\frac{d}{dt}Z_{ij}=-\frac{1}{8\pi^2}\left(\frac{1}{2}d_{kl}\lambda_{ikl}^*{Z_{km}^{-1}}^*{Z_{ln}^{-1}}^*\lambda_{jmn}-2C^{(r)}_{ij}Z_{ij}g_r^2\right), 
\eeq
where $d_{kl}$ and $C^r_{ij}$ are numerical factors appearing in the one-loop anomalous dimensions \cite{Evans:2013kxa,Martin:1993zk}, and $g_r$, ($r=1,2,3$) are running gauge couplings of $U(1)$, $SU(2)$ and $SU(3)$, respectively. As mentioned above, the initial condition for \eqref{RGEZ} is chosen to be $Z_{ij}(t_{GUT})=\delta_{ij}$. Beside the simplest case, it is hard to find an analytical solution of \eqref{RGEZ}. One of the possible ways out is to expand $Z_{ij}$ in power series:
%\footnote{We assume that $Z$ is analytical in the vicinity of $t_{GUT}$}
\beqa\label{serZ}
Z_{ij}(t)=\sum_{n=0}^{\infty}Z_{ij}^{(n)}(t_{GUT}).
\eeqa
One can show that to compute $Z^{(n)}_{ij}(t_{GUT})$, i.e. the $n$-th derivative of $Z_{ij}$ at $t_{GUT}$, it is enough to know all $Z^{(k)}_{ij}$ with $k<n$. As a result of such iterative procedure, $Z^{(n)}_{ij}$ can  be expressed in terms of small parameter $\epsilon=\ln 10/16\pi^2$, numerical factors $d_{kl}$ and $C_{ij}^{(r)}$, and values of holomorphic couplings $\lambda_{ijk}$, which match values of running physical couplings at $t_{GUT}$, and gauge couplings $g_r(t_{GUT})$  at the GUT scale. Two comments are in order here. Let us note that this method is especially useful for studying unification of gauge and Yukawa couplings because values of physical Yukawa couplings at low scale (e.g. the EWSB scale) can be expressed via their unified values, $y$ and $g_{GUT}$,  numerical factors $\epsilon$ and value of messenger scale $M$. 
The advantage of this method is that it allows to keep track how messenger-matter couplings enter values of Yukawa couplings at lower scales. An explicit example of described method is given in the next section. 

\begin{figure}
\begin{center}
\includegraphics[scale=0.6]{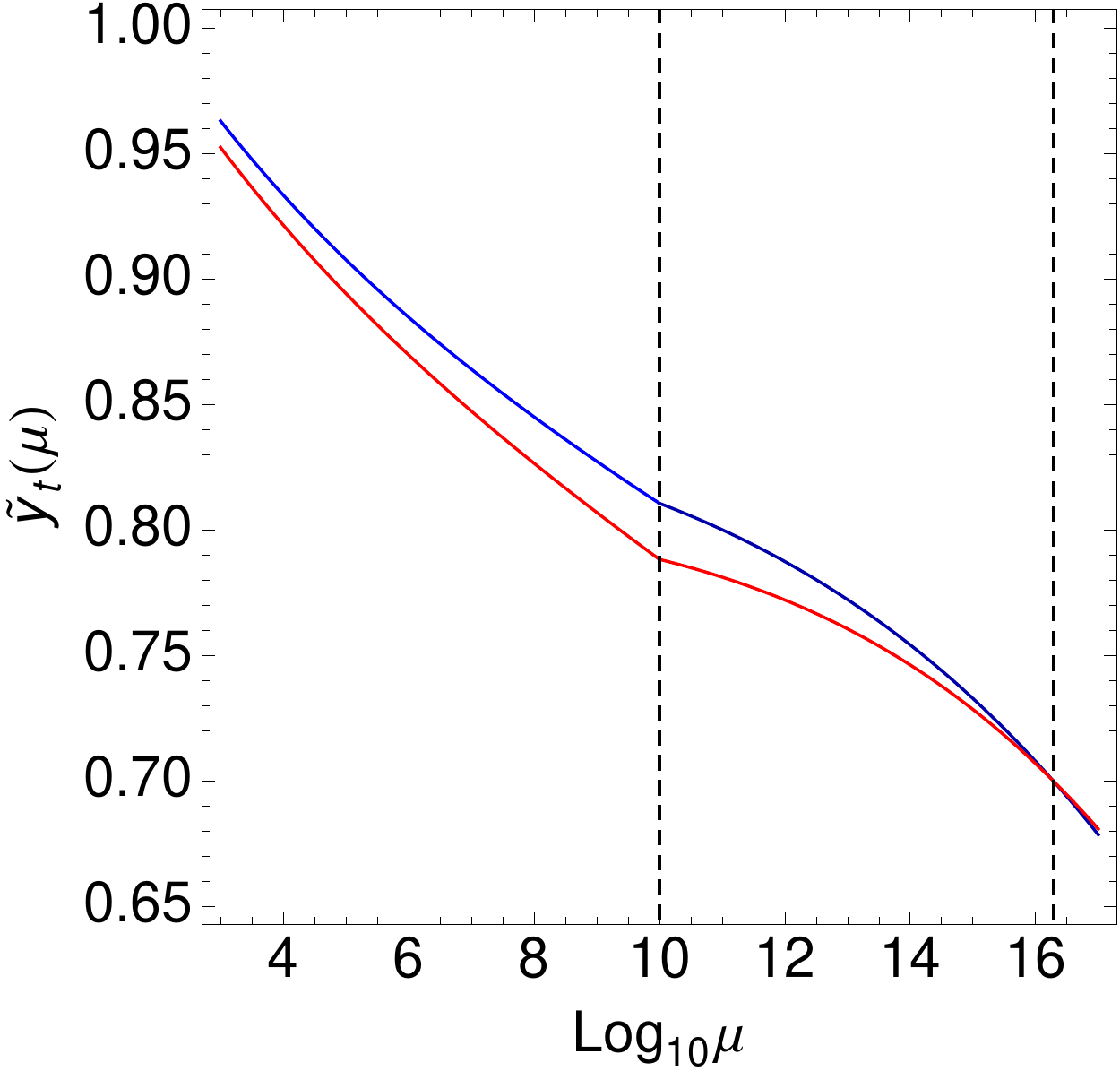}
\end{center}
\caption{Running of physical top Yukawa coupling $\widetilde{y}_t(\mu)$ when the mixing between messenger $Y_{\overline{U}}$ and MSSM field $\overline{U}$ is  
taken into account (blue) and when that effect is neglected (red). $\mu$ is renormalization scale in units of GeV. Messenger scale $M$ is set to $10^{10}\,\mathrm{GeV}$ (left vertical dashed line), while messenger-matter coupling is $h=0.4$. Value of top Yukawa coupling at GUT scale $M_{GUT}\approx10^{16}\,\mathrm{GeV}$ (right vertical dashed line) is set to $0.7$. Below the scale $M$, the top Yukawa coupling $\widetilde{y}_t$ runs according to the standard RG equation of the MSSM.}\label{ytcomp}
\end{figure}

It is instructive to compare
%see/follow/investigate/analyze 
%what is the difference 
%$\Delta y= \widetilde{y}-y_{RGE}$ 
%between 
the running of top Yukawa coupling $\widetilde{y}_t(t)$ when one neglects kinetic mixing above $M$
%between heavy and light states 
and evolves $\widetilde{y}_t(t)$ according to the standard RGE \cite{Martin:1993zk} with the running of $\widetilde{y}_t(t)$ derived from \eqref{lat}. To this end, let us consider the following superpotential with only one coupling, $H_uQY_{\overline{U}}$, between messenger $Y_{\overline{U}}$ and MSSM fields: 
\beq\label{Wex}
W=y_tH_uQ\overline{U}+h_tH_uQY_{\overline{U}}+MY_UY_{\overline{U}}. 
\eeq
Clearly, $Y_{\overline{U}}$ has the same charges as $\overline{U}$ and can mix with the latter.\footnote{To ensure gauge couplings unification, the spectrum has to contain also $Y_{Q}$, $Y_{\overline{E}}$ and their partners in conjugate representations. For simplicity, we assume that the superpotential couplings of those fields are suppressed with respect to the $H_uQY_{\overline{U}}$ coupling.} Here the messenger scale $M$ is set to $M=10^{10}\,\textrm{GeV}$, while values of the superpotential couplings at the GUT scale are the following: $y_t(t_{GUT})=y_t=0.7$, $h_t(t_{GUT})=h_t=0.4$. 
%
%We assume that the value of the top Yukawa at the GUT scale is $0.7$. 
Both solutions are presented on the Fig.~\ref{ytcomp}, where the blue curve shows  
%We can compare the running of top Yukawa coupling 
$\widetilde{y}_t(\mu)$
%, where $\mu$ is the renormalization scale, 
derived from  \eqref{lat}, 
%(blue curve on the Fig.~\ref{ytcomp}) 
while the red curve corresponds to 
%and 
the running obtained from standard RG equations without taking into account kinetic mixing. 
%between $Y_{\overline{U}}$ and $\overline{U}$ (red curve on the Fig.~\ref{ytcomp}). 
As one can see on Fig.~\ref{ytcomp}, the difference between the two values of $\widetilde{y}_t(\mu)$ grows when the renormalization scale $\mu$ decreases from $M_{GUT}$ to $M$, and then it becomes smaller when $\widetilde{y}_t(\mu)$ is evolved from $M$ down to the EWSB scale.  The discrepancy between both values of  top Yukawa coupling at the messenger scale $M$ is about 2\%. 
Let us stress that below the messenger scale Yukawa coupling $\widetilde{y}_t(\mu)$ runs according to the standard RGE of MSSM. If the kinetic mixing above $M$ 
is neglected then to get right value of $\widetilde{y}_t$ at the EWSB scale one has to compensate the deficit at $M$ 
%the threshold $M$/
%that scale one has to take into account %the threshold 
by a small correction 
%at $M$ i.e. 
which is precisely 
%one has to 
a shift  
%that value of Yukawa by $\Delta y_t=$ difference 
between the blue and red curve at the messenger scale $M$, see Fig.~\ref{ytcomp}. 

\section{An example of $SO(10)$ GUT model}\label{sec4}

Now, let us consider the following $SO(10)$ inspired GUT model. The spectrum of that model consists of one chiral field $H_{10}$ in representation $10$ of $SO(10)$, two chiral fields, $\phi_{16}$ and $Y_{16}$, in representation $16$, and one field, $Y_{\overline{16}}$, in representation $\overline{16}$.
We assume that below the GUT scale $M_{GUT}$, which is set by the condition $g_r(t_{GUT})=g_{GUT}$, the $SO(10)$ gauge symmetry is broken down to $SU(5)\times U(1)_{\chi}$ and further to $G_{SM}$. Under $SO(10)\to SU(5)\times U(1)_{\chi}$ the chiral fields decompose as follows:
\beq
%H_{10}:\quad
10\rightarrow5_2+\overline{5}_{-2},\quad\quad 
%\phi_{16}:\quad
16\rightarrow10_{-1}+\overline{5}_3+1_{-5},
\eeq
where subscripts denote $U(1)_{\chi}$ charges. Additionally, let us assume that after breaking of $SO(10)$ gauge symmetry, Higgs triplets and singlet components of $16$ and $\overline{16}$, $N_R$ and $Y_1$, receive masses of the order of GUT scale, and they decouple from the spectrum. It is straightforward to extend the analysis to the case when $N_R$ and $Y_1$ are lighter than $M_{GUT}$. The superpotential of the discussed model is of the form:
\beq\label{WSO10}
W=
yH_{10}\phi_{16}\phi_{16}+hH_{10}\phi_{16}Y_{16}+MY_{16}Y_{\overline{16}}.
\eeq
Let us note that two other messenger-matter couplings are also possible, $H_{10}Y_{16}Y_{16}$ and $H_{10}Y_{\overline{16}}Y_{\overline{16}}$, but, for the simplicity, we assume that they are suppressed with respect to those in \eqref{WSO10}. 
\begin{figure}
\begin{center}
\includegraphics[scale=0.6]{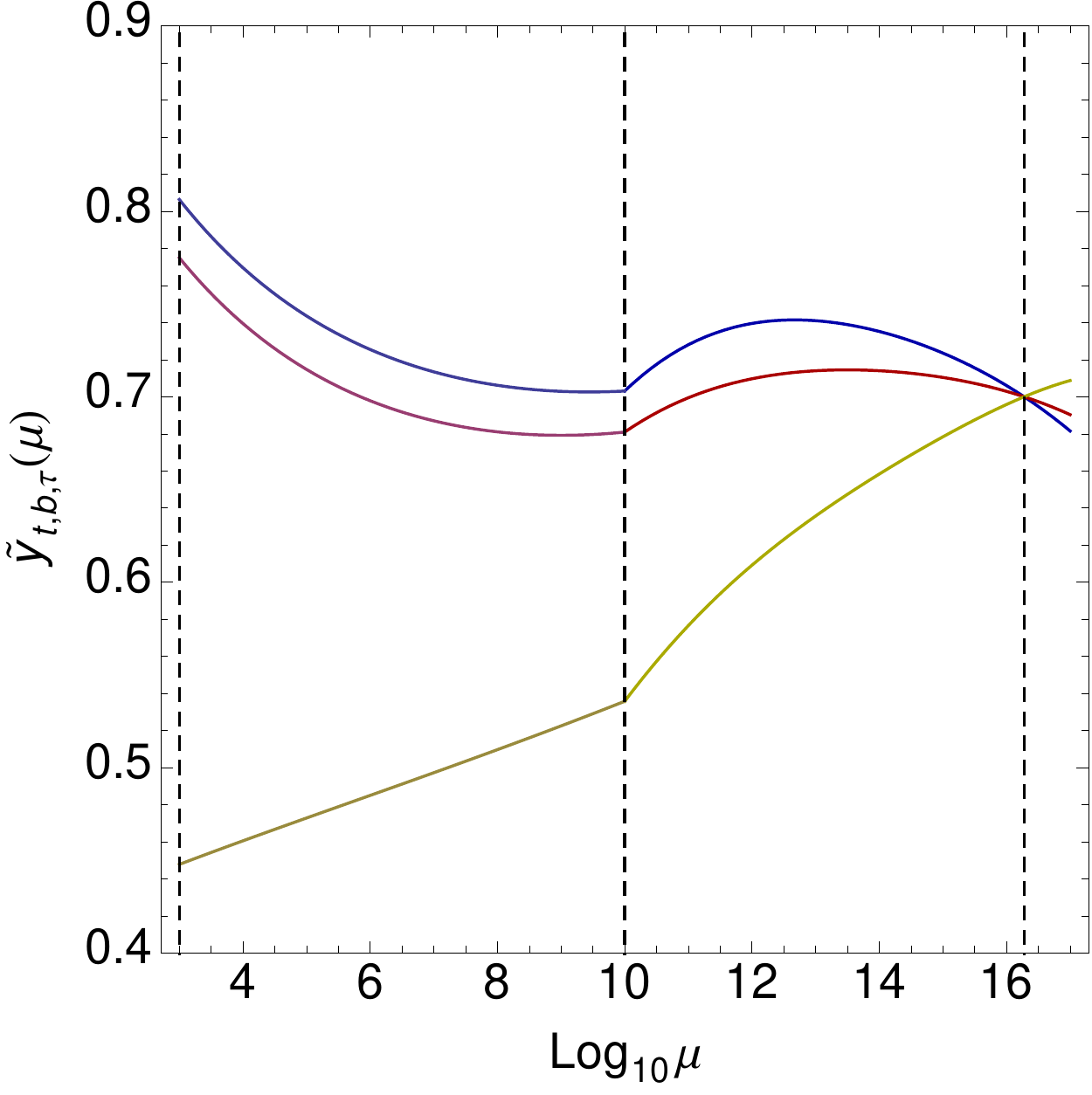}
\end{center}
\caption{Running of top (blue curve), bottom (magenta curve) and tau (yellow curve) Yukawa couplings $\widetilde{y}_{t,b,\tau}(\mu)$ in the $SO(10)$ inspired GUT model discussed in the Sec. 4. $\mu$ is the renormalization scale in $\mathrm{GeV}$ units. Messenger scale is set to $10^{10}\,\mathrm{GeV}$ (middle vertical dashed line). Unified value of Yukawa couplings is set to $y_{GUT}=0.7$ while messenger-matter coupling is $h=0.4$. Left vertical dashed line corresponds to the EWSB scale, while the right vertical dashed line shows the GUT scale.}\label{ySO10}
\end{figure}
We applied the method presented in Sec.~\ref{sec3} to survey the issue of top-bottom-tau unification in the following way. First, the precise unification of Yukawa couplings was assumed:  $y=y_t(t_{GUT})=y_b(t_{GUT})=y_{\tau}(t_{GUT})$. 
%Parameters of the model are the following: $y=0.7$, $h=0.4$, $M=10^{10}\,\textrm{GeV}$.
Then wave-function renormalization matrix $Z_{ij}(t)$ was derived from \eqref{RGEZ} and \eqref{serZ}. As an example,  
%For the completness, 
we present an explicit expression for $Z_{H_uH_u}(t)$ at scale $t=\ln\mu/Q$ in the discussed model:
% derived using \eqref{RGEZ} and \eqref{serZ}: 
\begin{eqnarray}\label{ZHuHu}
Z_{H_uH_u}(t)&=&1+\frac{6}{5}\epsilon[3g_{GUT}^2-5(2h^2-y^2)](t-t_{GUT})\nn\\
&&+\frac{24}{25}\epsilon^2[29g_{GUT}^2+35(2h^2+y^2)-25(2h^4+4h^2y^2+y^4)](t-t_{GUT})^2+\ldots
\end{eqnarray}
As mentioned in Sec.~\ref{sec3}, 
%when precise unification is assumed then the wave-function renormalization matrix $Z$ at scale $t=\ln\mu/Q$ 
$Z_{ij}(t)$ can be expressed in terms of small parameter $\epsilon$ and GUT parameters of the model: $t_{GUT}$, $y$ and $h$. Finally, the values of physical Yukawa couplings $\widetilde{y}_{t,b,\tau}(t)$ at the messenger scale $M$ were obtained from \eqref{lambdaphys}. 
In the next step, derived values of Yukawa couplings $\widetilde{y}_{t,b,\tau}(t)$ were appropriately implemented in the code of 
%\texttt{SPheno} [Ref] and 
\texttt{SuSpect} \cite{Djouadi:2002ze}. Morover, we computed 1- and 2-loop soft terms generated by \eqref{WSO10}, and also implemented them into the code of 
%\texttt{SPheno} and 
\texttt{SuSpect}. Then we used these numerical routines
% software 
to derive physical spectrum at the EWSB scale and compared obtained values of $\widetilde{y}_{t,b,\tau}(t)$ at the EWSB scale $t_{EWSB}$ with the values obtained from $t$, $b$ and $\tau$ pole masses. An example of the running of $\widetilde{y}_{t,b,\tau}(t)$ is shown on Fig.~\ref{ySO10}.
%with the help of appropirately modified codes of SPheno [ref] and SuSpect [ref]. 
We scanned over the following range of parameters: $8<t_M=\log_{10}M<14$, $0.6<y<0.9$ and $0<h<1.2$ and checked if the simplest low-energy constraints are satisfied i.e.: (a) whether lightest neutral Higgs mass is about $125\,\textrm{GeV}$, (b) gluino $\widetilde{g}$ and 1st and 2nd generation squarks $\widetilde{q}_{1,2}$ masses  are bigger than $1.8\,\textrm{TeV}$ and, finally, (c) whether EWSB vacuum is stable.  
%(i.e. there is no charge or colour breaking and $\mathrm{UFB/CCB}$). 
We obtained the following results. When $\widetilde{y}_t(t_{EWSB})$ and $\widetilde{y}_{\tau}(t_{EWSB})$ match corresponding values derived from pole masses of $t$ and $\tau$, then one can get about $20\%$ discrepancy between value of $\widetilde{y}_b(t_{EWSB})$ and the related value $y_b^{(0)}$ derived from pole mass of $b$ if: (i) $\tan\beta$ is about $45$ and (ii) one allows for tachyonic $\widetilde{\tau}$. On the other hand, when $\tan\beta$ is smaller, about $20$, then there are no tachyons in the spectrum, but there is also large mismatch, of the order of factor 2, between $\widetilde{y}_{b}(t_{EWSB})$ and $y^{(0)}_b$. 
 %analyzed phenomenology and compare values of Yukawa couplings derived from the presented method with values computed from fermions masses. 
%
Hence in the discussed minimal $SO(10)$ model, it is not possible to reconcile precise top-bottom-tau unification with the low-energy constraints on the spectrum.   
To avoid instability of the potential related to $\widetilde{\tau}$ tachyonic mass, one could, for example, extend spectrum or allow additional messenger couplings. 

\section{Summary}

We have discussed the issue of Yukawa couplings renormalization in the presence of messenger-matter interactions, which occur in the EGMSB models. Such messenger-matter couplings not only generate 1- and 2-loop soft terms but can also lead to kinetic mixing between heavy and light fields. It turns out that this   %kinetic 
mixing 
%induced by mess-matt couplings 
results in small shifts of the values of Yukawa couplings at the messenger scale, what in turn influence their values 
%of Yukawa couplings 
at the EWSB scale.  We presented a method based on RGE for wave-function renormalization and showed that it is a handy tool to analyse RG flow of Yukawa coupling, especially useful in the top-bottom approach.
% this method can be implemented in a similar way at 2-loop level.

We applied proposed method to the specific $SO(10)$ inspired GUT model and showed that in this scenario it is not possible to reconcile precise top-bottom-tau unification with  non-tachyonic spectrum, stable EWSB vacuum and right values of Yukawa couplings at the EWSB scale. It turned out that phenomenology of the discussed model was spoiled by tachyonic  $\widetilde{\tau}$.  One of the possible ways out is to extend spectrum of the model or allow for additional messenger-matter couplings. 
Although the discussed example is quite specific, the presented method is general and can be used for studying Yukawa couplings renormalization in other, maybe less rigid, GUT models as $SU(5)$, or other $SO(10)$ scenarios which allow for small deviations from precise top-bottom-tau unification. 
% can be also tested and analyze e.g. $b-\tau$ unification issue. 
%more messenger-matter couplings are switched

\acknowledgments
The author would like to thank Marcin Badziak for useful discussions and comments and to Jared Evans for remarks on the method presented in \cite{Evans:2013kxa}. This work was supported by the Polish National Science Centre (NCN) under postdoctoral grant No. DEC-2012/04/S/ST2/00003 and in part by the Institute of Physics,  University of Silesia under Young Scientists Grant 2013.


\begin{thebibliography}{99}

\bibitem{Giudice:1998bp}
  G.~F.~Giudice and R.~Rattazzi,
  \emph{Theories with gauge mediated supersymmetry breaking},
  \emph{Phys.\ Rept.}\  {\bf 322} (1999) 419
  [{\tt hep-ph/9801271}].
  %%CITATION = HEP-PH/9801271;%%
  %1276 citations counted in INSPIRE as of 29 mar 2015
  
\bibitem{Draper:2011aa}
  P.~Draper, P.~Meade, M.~Reece and D.~Shih,
  \emph{Implications of a 125 GeV Higgs for the MSSM and Low-Scale SUSY Breaking},
  \emph{Phys.\ Rev.\ D} {\bf 85} (2012) 095007
  [{\tt arXiv:1112.3068 [hep-ph]}].
  %%CITATION = ARXIV:1112.3068;%%
  %228 citations counted in INSPIRE as of 24 Apr 2015   
  
\bibitem{Dine:1996xk}
  M.~Dine, Y.~Nir and Y.~Shirman,
  \emph{Variations on minimal gauge mediated supersymmetry breaking},
  \emph{Phys.\ Rev.\ D} {\bf 55} (1997) 1501
  [{\tt hep-ph/9607397}].
  %%CITATION = HEP-PH/9607397;%%
  %196 citations counted in INSPIRE as of 29 mar 2015
  
\bibitem{Giudice:1997ni}
  G.~F.~Giudice and R.~Rattazzi,
  \emph{Extracting supersymmetry breaking effects from wave function renormalization},
  \emph{Nucl.\ Phys.\ B} {\bf 511} (1998) 25
  [{\tt hep-ph/9706540}].
  %%CITATION = HEP-PH/9706540;%%
  %264 citations counted in INSPIRE as of 29 mar 2015  



\bibitem{Chacko:2001km}
  Z.~Chacko and E.~Ponton,
  \emph{Yukawa deflected gauge mediation},
  \emph{Phys.\ Rev.\ D} {\bf 66} (2002) 095004
  [{\tt hep-ph/0112190}].
  %%CITATION = HEP-PH/0112190;%%
  %63 citations counted in INSPIRE as of 29 Mar 2015

\bibitem{Joaquim:2006uz}
  F.~R.~Joaquim and A.~Rossi,
  \emph{Gauge and Yukawa mediated supersymmetry breaking in the triplet seesaw scenario},
  \emph{Phys.\ Rev.\ Lett.}  {\bf 97} (2006) 181801
  [hep-ph/0604083].
  %%CITATION = HEP-PH/0604083;%%

\bibitem{Joaquim:2006mn}
  F.~R.~Joaquim and A.~Rossi,
  \emph{Phenomenology of the triplet seesaw mechanism with Gauge and Yukawa mediation of SUSY breaking},
  \emph{Nucl.\ Phys.\ B} {\bf 765} (2007) 71
  [{\tt hep-ph/0607298}].
  %%CITATION = HEP-PH/0607298;%%
  %56 citations counted in INSPIRE as of 30 Mar 2015  

\bibitem{Evans:2013kxa}
  J.~A.~Evans and D.~Shih,
  \emph{Surveying Extended GMSB Models with $m$$_{h}$=125 GeV},
  \emph{JHEP} {\bf 1308} (2013) 093
  [{\tt arXiv:1303.0228 [hep-ph]}].
  %%CITATION = ARXIV:1303.0228;%%
  %23 citations counted in INSPIRE as of 29 Mar 2015

\bibitem{Jelinski:2013kta}
  T.~Jeli\'nski,
  \emph{On messengers couplings in extended GMSB models},
  \emph{JHEP} {\bf 1309} (2013) 107
  [{\tt arXiv:1305.6277 [hep-ph]}].
  %%CITATION = ARXIV:1305.6277;%%
  %4 citations counted in INSPIRE as of 29 Mar 2015  



  \bibitem{Evans:2011bea}
  J.~L.~Evans, M.~Ibe and T.~T.~Yanagida,
  \emph{Relatively Heavy Higgs Boson in More Generic Gauge Mediation},
  \emph{Phys.\ Lett.\ B} {\bf 705} (2011) 342
  [arXiv:1107.3006 [hep-ph]].
  %%CITATION = ARXIV:1107.3006;%%
  %44 citations counted in INSPIRE as of 24 Apr 2015

\bibitem{Jelinski:2011xe}
  T.~Jelinski, J.~Pawelczyk and K.~Turzynski,
 \emph{On Low-Energy Predictions of Unification Models Inspired by F-theory},
  \emph{Phys.\ Lett.\ B} {\bf 711} (2012) 307
  [arXiv:1111.6492 [hep-ph]].
  %%CITATION = ARXIV:1111.6492;%%
  %8 citations counted in INSPIRE as of 24 Apr 2015
  
\bibitem{Kang:2012ra}
  Z.~Kang, T.~Li, T.~Liu, C.~Tong and J.~M.~Yang,
  \emph{A Heavy SM-like Higgs and a Light Stop from Yukawa-Deflected Gauge Mediation},
  \emph{Phys.\ Rev.\ D} {\bf 86} (2012) 095020
  [arXiv:1203.2336 [hep-ph]].
  %%CITATION = ARXIV:1203.2336;%%
  %56 citations counted in INSPIRE as of 24 Apr 2015 

\bibitem{Albaid:2012qk}
  A.~Albaid and K.~S.~Babu,
  \emph{Higgs boson of mass 125 GeV in GMSB models with messenger-matter mixing},
  \emph{Phys.\ Rev.\ D} {\bf 88} (2013) 055007
  [arXiv:1207.1014 [hep-ph]].
  %%CITATION = ARXIV:1207.1014;%%

\bibitem{Shadmi:2011hs}
  Y.~Shadmi and P.~Z.~Szabo,
  \emph{Flavored Gauge-Mediation},
  \emph{JHEP} {\bf 1206} (2012) 124
  [arXiv:1103.0292 [hep-ph]].
  %%CITATION = ARXIV:1103.0292;%%
  
  \bibitem{Abdullah:2012tq}
  M.~Abdullah, I.~Galon, Y.~Shadmi and Y.~Shirman,
  \emph{Flavored Gauge Mediation, A Heavy Higgs, and Supersymmetric Alignment},
  \emph{JHEP} {\bf 1306} (2013) 057
  [arXiv:1209.4904 [hep-ph]].
  %%CITATION = ARXIV:1209.4904;%%

\bibitem{Calibbi:2013mka}
  L.~Calibbi, P.~Paradisi and R.~Ziegler,
  \emph{Gauge Mediation beyond Minimal Flavor Violation},
  \emph{JHEP} {\bf 1306} (2013) 052
  [arXiv:1304.1453 [hep-ph]].
  %%CITATION = ARXIV:1304.1453;%%

\bibitem{Galon:2013jba}
  I.~Galon, G.~Perez and Y.~Shadmi,
  \emph{Non-Degenerate Squarks from Flavored Gauge Mediation},
  JHEP {\bf 1309} (2013) 117
  [arXiv:1306.6631 [hep-ph]].
  %%CITATION = ARXIV:1306.6631;%%
  
\bibitem{Jelinski:2014uba}
  T.~Jelinski and J.~Pawelczyk,
  \emph{Masses and FCNC in Flavoured GMSB scheme},
  arXiv:1406.4001 [hep-ph].
  %%CITATION = ARXIV:1406.4001;%%
  %2 citations counted in INSPIRE as of 24 Apr 2015  

\bibitem{Yukawa-HV1}
  J.~J.~Heckman and C.~Vafa,
  \emph{Flavor Hierarchy From F-theory},
  \emph{Nucl.\ Phys.\ B} {\bf 837} (2010) 137
  [arXiv:0811.2417 [hep-th]].
  %%CITATION = ARXIV:0811.2417;%%

\bibitem{HV-E8}
  J.~J.~Heckman, A.~Tavanfar and C.~Vafa,
  \emph{The Point of E(8) in F-theory GUTs},
  \emph{JHEP} {\bf 1008} (2010) 040
  [arXiv:0906.0581 [hep-th]].
  %%CITATION = ARXIV:0906.0581;%%
	
\bibitem{Pawelczyk:2010xh}
  J.~Pawelczyk,
  \emph{F-theory inspired GUTs with extra charged matter},
  \emph{Phys.\ Lett.\ B} {\bf 697} (2011) 75
  [arXiv:1008.2254 [hep-ph]].
  %%CITATION = ARXIV:1008.2254;%%
  %9 citations counted in INSPIRE as of 24 Apr 2015	

\bibitem{Froggatt-Nielsen}
  C.~D.~Froggatt and H.~B.~Nielsen,
  \emph{Hierarchy of Quark Masses, Cabibbo Angles and CP Violation},
  \emph{Nucl.\ Phys.\ B} {\bf 147} (1979) 277.
  %%CITATION = NUPHA,B147,277;%%

\bibitem{Bagger:1996ei}
  J.~A.~Bagger, K.~T.~Matchev, D.~M.~Pierce and R.~J.~Zhang,
  \emph{Gauge and Yukawa unification in models with gauge mediated supersymmetry breaking},
  \emph{Phys.\ Rev.\ Lett.}  {\bf 78} (1997) 1002
   [\emph{Phys.\ Rev.\ Lett.}  {\bf 78} (1997) 2497]
  [{\tt hep-ph/9611229}].
  %%CITATION = HEP-PH/9611229;%%
  %47 citations counted in INSPIRE as of 24 Apr 2015  

\bibitem{Pierce:1996zz}
  D.~M.~Pierce, J.~A.~Bagger, K.~T.~Matchev and R.~j.~Zhang,
  \emph{Precision corrections in the minimal supersymmetric standard model},
  \emph{Nucl.\ Phys.\ B} {\bf 491} (1997) 3
  [hep-ph/9606211].
  %%CITATION = HEP-PH/9606211;%%
  %871 citations counted in INSPIRE as of 24 Apr 2015 

 \bibitem{Blazek:2002ta}
  T.~Blazek, R.~Dermisek and S.~Raby,
  \emph{Yukawa unification in SO(10)},
  \emph{Phys.\ Rev.\ D} {\bf 65} (2002) 115004
  [hep-ph/0201081].
  %%CITATION = HEP-PH/0201081;%%
  %171 citations counted in INSPIRE as of 24 Apr 2015
 
\bibitem{Raby:2003in}
  S.~Raby,
  \emph{Phenomenology of the minimal SO(10) SUSY model},
  \emph{Pramana} {\bf 62} (2004) 523
  [hep-ph/0304074].
  %%CITATION = HEP-PH/0304074;%%
  %3 citations counted in INSPIRE as of 24 Apr 2015 
 
\bibitem{Badziak:2011wm}
  M.~Badziak, M.~Olechowski and S.~Pokorski,
  \emph{Yukawa unification in SO(10) with light sparticle spectrum},
  \emph{JHEP} {\bf 1108} (2011) 147
  [arXiv:1107.2764 [hep-ph]].
  %%CITATION = ARXIV:1107.2764;%%
  %24 citations counted in INSPIRE as of 24 Apr 2015

\bibitem{Byakti:2013ti}
  P.~Byakti and T.~S.~Ray,
  \emph{Burgeoning the Higgs mass to 125 GeV through messenger-matter interactions in GMSB models},
  \emph{JHEP} {\bf 1305} (2013) 055
  [{\tt arXiv:1301.7605 [hep-ph]}].
  %%CITATION = ARXIV:1301.7605;%%
  %14 citations counted in INSPIRE as of 29 mar 2015

\bibitem{Craig:2012xp}
  N.~Craig, S.~Knapen, D.~Shih and Y.~Zhao,
  \emph{A Complete Model of Low-Scale Gauge Mediation},
  \emph{JHEP} {\bf 1303} (2013) 154
  [{\tt arXiv:1206.4086 [hep-ph]}].
  %%CITATION = ARXIV:1206.4086;%%
  %52 citations counted in INSPIRE as of 24 Apr 2015   

 \bibitem{Grisaru:1979wc}
  M.~T.~Grisaru, W.~Siegel and M.~Rocek,
 \emph{Improved Methods for Supergraphs},
  \emph{Nucl.\ Phys.\ B} {\bf 159} (1979) 429.
  %%CITATION = NUPHA,B159,429;%%
  %818 citations counted in INSPIRE as of 24 Apr 2015

\bibitem{Horn}
R.~A.~Horn and C.~R.~Johnson, \emph{Matrix Analysis}, CUP, New York, 1990.

\bibitem{Martin:1993zk}
  S.~P.~Martin and M.~T.~Vaughn,
  \emph{Two loop renormalization group equations for soft supersymmetry breaking couplings},
  \emph{Phys.\ Rev.\ D} {\bf 50} (1994) 2282
   [Erratum-ibid.\ \emph{D} {\bf 78} (2008) 039903]
  [{\tt hep-ph/9311340}].
  %%CITATION = HEP-PH/9311340;%%
  %640 citations counted in INSPIRE as of 30 Mar 2015  

\bibitem{Djouadi:2002ze}
  A.~Djouadi, J.~-L.~Kneur and G.~Moultaka,
  \emph{SuSpect: A Fortran code for the supersymmetric and Higgs particle spectrum in the MSSM},
  \emph{Comput.\ Phys.\ Commun.}  {\bf 176} (2007) 426 [hep-ph/0211331].
  

\end{thebibliography}
\end{document}